# The effectiveness of data augmentation in porous substrate, nanowire, fiber and tip images at the level of deep learning intelligence

[1]C.H.Wong[*], [2]S.M. Ng, [2]C.W.Leung, [1]A.F.Zatsepin

[1]Institute of Physics and Technology, Ural Federal University, Yekaterinburg, Russia

[2]Department of Applied Physics, The Hong Kong Polytechnic University, Hong Kong, China

Email: ch.kh.vong@urfu.ru

**Abstract:**

To prepare for identifying the composition of nanowire-fiber mixtures in Scanning Electron Microscope (SEM) images, we optimize the performance of image classification between nanowires, fibers and tips due to their geometric similarities. The SEM images are analyzed by deep learning techniques where the validation accuracies of 11 convolutional neural network (CNN) models are compared. By increasing the diversity of data such as reflection, translation and scale factor approaches, the highest validation accuracy of recognizing nanowires, fibers and tips is 97.1%. We proceed to classify the level of porosity in anodized aluminum oxide for the self-assisted nanowire growth where the validation accuracy is optimized at 93%. Our software allow scientists to count the percentage of fibers in any nanowire-fiber composite and design the porous substrate for embedding different sizes of nanowires automatically, which assists the software development in Nanoscience Foundries & Fine Analysis (NFFA) Europe Projects.

**Introduction:**

Image recognition is one of the leading computer vision technologies that allows the digital world interacting with the physical world [1-5]. Computer vision engineers always label images into different categories for image classification. In the last decade, image recognition was widely implemented by machine learning algorithms such as the Support Vector Machine and the Cascade Object Detector [5]. Machine learning algorithms analyze data and learn from those data to announce informed decisions only [6]. In contrast, deep learning models intensify the algorithms layer-by-layer to establish an artificial neural network where the software may learn and make wise decisions on its own [7].

Deep learning is an evolution of artificial intelligence. The convolutional neural network (CNN) is a type of deep neural networks [8] for artificial intelligence. As a rule of thumb, establishing a pre-training network requires at least ~500 images per category. The architecture of CNN involves image input layer, convolutional layer, batch normalization layer, ReLU layer, cross channel normalization layer, pooling layers, dropout Layer, fully connected layer and output layer [9]. When deep neural networks combine with data augmentation, it is always effective to improve the success rate of image recognition [10]. Data augmentation utilizes a domain-specific matrix to mimic the existing data to generate 'new' training samples artificially [11,12]. However, there is no universal rule to set up the parameters of data augmentation. Empirical methods are always used to find out the best method of data-augmentation for a particular category [10-13].

Scanning electron microscopes (SEM) is a powerful machine to observe the structure of materials in nanoscale [14]. However, human eyes always struggle to classify SEM images unless the researchers spend a tremendous amount of time to analyze them. This is the origin of the Nanoscience Foundries & Fine Analysis (NFFA) Europe Projects [15] which is collaborated by 20 European nanoscience research laboratories that aims at reorganizing SEM images under an artificial intelligence platform such as the cellular organisms, the dendrites and the morphology of carbon nanotubes…etc. [16,17].

Nanowire-Fiber composite is a valuable research topic in materials science. The silver-nanowire-cellulose-fiber composite can improve the sensitivity of detecting pressure and proximity [18]. In addition, the $Si_3N_4$ nanowire-carbon nanotube networks show a strong impact on aerospace engineering [19]. However, human eyes are not effective to recognize the nanowires and fibers where their geometric similarities can be observed in Figure 1a & 1c [14]. Worse still, it is an uphill struggle to count the percentage of nanowires in any nanowire-fiber composite because the SEM image contains nanowires and fibers simultaneously. Hence, our first project is to improve the image recognition between nanowires and fibers [14] with the help of data augmentation. After the best method of augmenting these two subjects is found, it will be easier for scientists to examine the percentage of nanowires in any nanowire-fiber composite. As the shape of an isolated nanowire or fiber is similar to tip, we add it into the SEM dataset where their geometric similarities can be observed in Figure 1b & 1d.

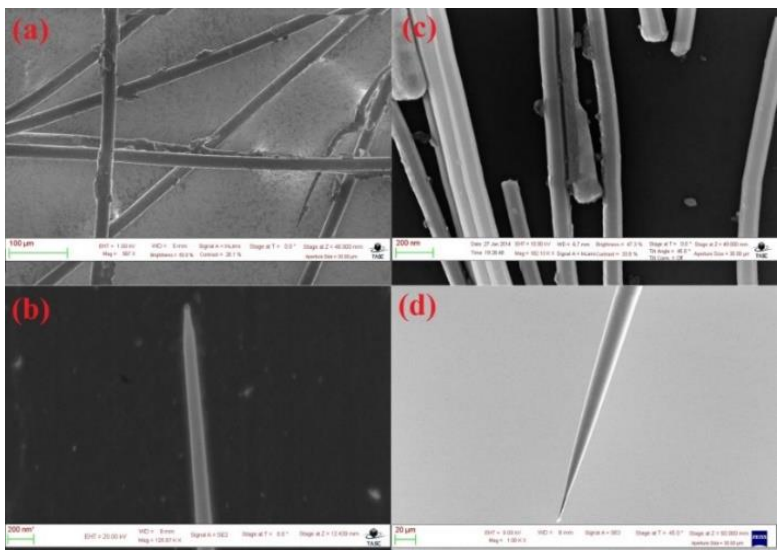

Figure 1: The SEM images of (a) fibers, (b) an isolated nanowire (c) nanowires and (d) a tip [14].

Another valuable object for image classification is the porosity of anodized aluminium oxide (AAO). Controlling the porosity of the AAO substrate is advantageous to prepare nanoparticles, nanowires and nanotubes. The dimensions of the AAO holes (pore diameter and pore-to-pore distance) depend on the applied voltage where the geometric similarities are displayed in Figure 2 [20,21]. It should be noted that the AAO as shown in Figure 2 was prepared by soft ultraviolet nanoimprint pre-patterning of aluminum films (as a guide of pore growth) followed by anodization in phosphoric acid with different voltages. However, embedding nanowires in different diameters require dissimilar sizes of the AAO holes. When the size of the AAO holes varies as a function of space, the image classification of the AAO holes assists scientists to examine whether the distribution of the dissimilar AAO holes is desired or not. Hence, our second project is to find out the best method of augmenting the AAO images in order to prepare for the AAO substrate for embedding different types of nanowires in near future.

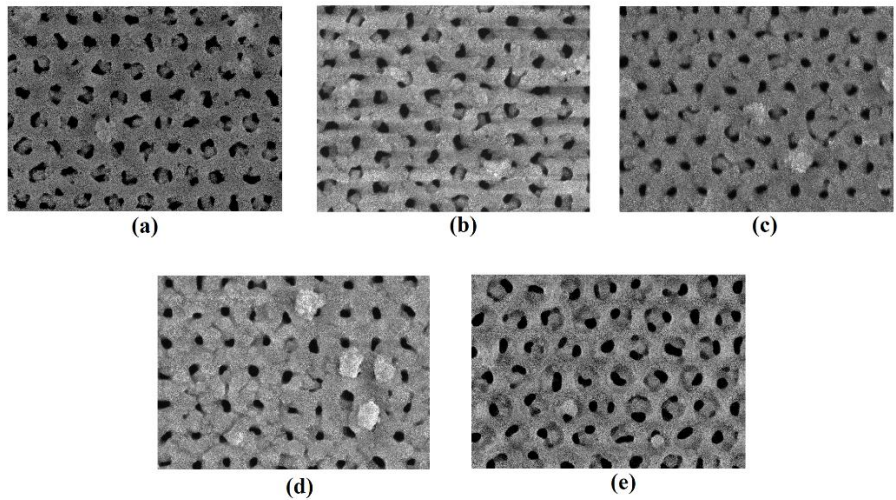

Figure 2: The porosity of anodized aluminum oxide prepared with different voltages (a) 110V (b) 130V (c) 150V (d) 170V (e) 190V [20,21].

**Computational methods:**

We used a FUJITSU computer equipped with Intel(R) Core(TM) i7-3632QM CPU at 2.2GHz to implement deep learning tasks. The SEM images of nanowires, fibers and tips can be downloaded in ref 14. These three categories, "Nanowire", "Fiber" and "Tip" [14], are abbreviated as "NFT" objects where 152 images per category is assigned. The definition of "Nanowire" and "Fiber" can be found in reference 14. The AAO images are provided by the Department of Applied Physics at The Hong Kong Polytechnic University [20,21]. There are five image categories ("110V", "130V", "150V, "170V", "190V") in the AAO dataset where 20 images per category are randomly chosen for training. We use the famous SqueezeNet, ShuffleNet, GoogLeNet, MobileNetV2, ResNet-18, Darknet-19, ResNet-50, NASNet-Mobile,

ResNet-101, DenseNet and Xception networks to train the SEM dataset, respectively [22]. Eleven validation accuracies are compared in which the data augmentation is only applied to the top-score network. Data augmentation increases the amount of training data but it may reduce overfitting on computational models [10-13]. The flowchart of the deep learning process is drawn in Figure 3. We aim at evaluating the model's performance under different methods of data augmentation and therefore monitoring validation accuracy is already representative. The data augmentations are applied in the following ways [10-13].

(1) Random reflection along X direction under a specified scalar. The reflection factor is always activated unless otherwise specified.

(2) The angle of rotation is chosen randomly from a continuous uniform distribution within the specified interval $[-\theta_R, \theta_R]$. The maximum rotational angle is abbreviated as $\theta_R$.

(3) The scale factor is assigned in a random manner from a continuous uniform distribution within the specified interval $[1-S_F, 1+S_F]$ along the XY plane. The maximum scale factor is labelled as $S_F$. For instance, $S_F$ equals to 10% which is equivalent to the specified interval of [0.9,1.1].

(4) The shear angles along Y direction are assigned randomly from a continuous uniform distribution within the maximum shear angle $[-\theta_S, \theta_S]$.

(5) The translational widths are selected randomly from a continuous uniform distribution within the maximum translational interval [-T,T] along X and Y direction, respectively.

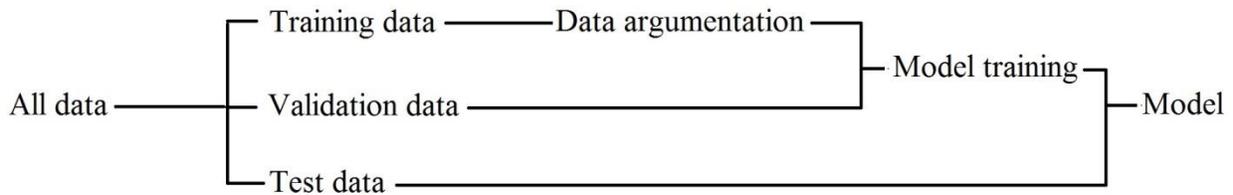

Figure 3: The flowchart of the deep learning process. We split the data into 70% training and 30% validation.

**Results and Discussions:**

We display the validation accuracies in different neural networks in Figure 4. The computational cost of SqueezeNet is the smallest but its validation accuracy is less than 90%. The GoogLeNet shows the highest accuracies at 92.7% in a reasonable computational cost. The poorest neural network for recognizing the "NFT" objects is ResNet-18. Using the other neural networks with higher computational costs does not show further improvement in terms of validation accuracy. The GoogLeNet [24] makes use of the inception modules and allows the network to choose different sizes of convolutional filters in each block [22]. The Inception network stacks these modules layer-by-layer with occasional max-pooling layers to optimize the resolution of the grids [22]. On the other hand, the depth of the layer plays an important role to the ResNet family. ResNet-18, ResNet-50 and ResNet-100 are the convolutional neural networks that contain 18, 50 and 100 layers, respectively [23]. The more layers the ResNet involves, the higher validation accuracy it attains [22,23].

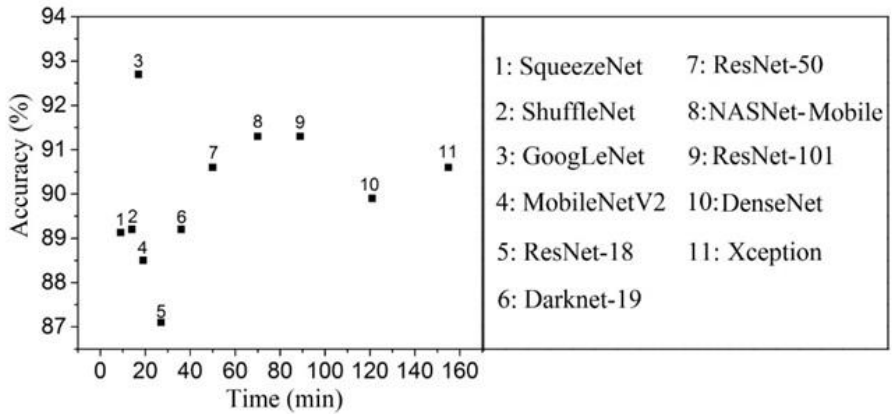

Figure 4: The validation accuracies of classifying the "NFT" objects in various deep learning algorithms.

We have found that the GoogLeNet [24] is very accurate to identify the "NFT" objects where the validation accuracy without data argumentation reaches 92.7% already. In order to improve the validation accuracy even better, we search for the desired matrices to augment [10-13] the SEM image data. Figure 5 illustrates that the translational matrix influences the validation accuracy significantly. The translation distance is picked in a random manner from a continuous uniform distribution within the specified interval [-T,T]. When the ±T interval is selected at [-10,10], the validation accuracy increases from 92.7% to 93.4%. Setting the ±T interval to [-20,20] makes the validation accuracy slightly better only. The best ±T interval to classify the "NFT" objects is [-30,30] where the validation accuracy is as high as 94.9%. However, the ±T intervals at [-40,40] and [-50,50] do not enhance the image classification anymore. Symmetry gives significant impacts to image classification [25]. When the SEM users capture images, they always try to locate the target at the best position. However, the target may not be perfectly located due to human errors. The random shift within the translational intervals likely remedy the loss of feature detection due to the human errors. On the other hand, the translational intervals beyond the critical limit reduce the success rate of image classification because the main features in the images may become unreadable.

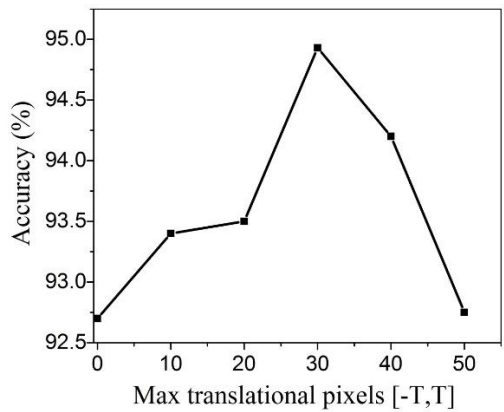

Figure 5: The validation accuracy of the "NFT" objects is affected by translational augmentation. The image dataset is randomly shifted within the ±T interval.

Using translational matrix [28] is not the only way to augment image data. Scaling images [26,28] is another method to improve the validation accuracy. We fix the ±T interval to [-30,30] and then search for the optimal scale factor. The maximum scale factor $S_F$ is applied to the XY plane simultaneously. Figure 6 confirms that scaling the images within ±10% raises the validation accuracy to 97.1%. Increasing the scale factor to ±20% does not enhance the validation accuracy. Setting the maximum scale factor to ±30% drops the validation accuracy. Scaling images may emerge the features predominantly. For example, magnifying the objects with a reasonable magnification power can visualize the features more easily. A wrong magnification power destroys the performance of image recognition. If the scale factor is too large, the images become blurry and therefore the validation accuracy decreases. In contrast, diminishing the images may hide the poorly captured regions that prevents the occurrence of a poor image classification.

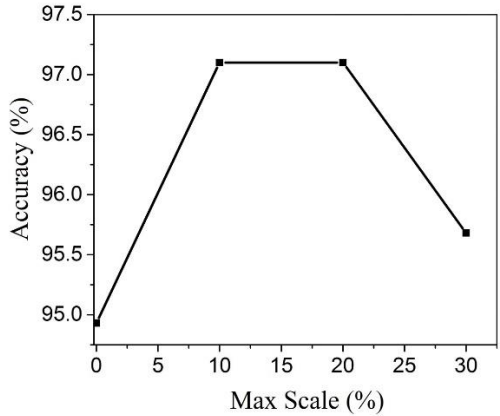

Figure 6: The validation accuracy of the "NFT" objects is influenced by scale-factor augmentation.

By fixing the ±T interval to [-30,30] and the ±$S_F$ interval to 10%, we examine whether the rotational and shear augmentations improve the validation accuracy or not [27,28]. According to Figure 7, rotating the SEM images within the ±$\theta_R$ interval of [-45°,45°] reduces the validation accuracy to 92.6%. Widening the ±$\theta_R$ interval to [-90°,90°] makes the image classification very poor. Extending the ±$\theta_R$ interval to [-135°,135°] and [-180°,180°] restores the validation accuracy above 90% again. The V-like shape in Figure 7 is owing to the loss of vertical symmetry during rotation. For the "NFT" dataset, most nanowires, fibers and tips propagate along the vertical direction [14]. The use of ±$\theta_R$ = ±90° points the propagation axis toward the horizontal axis and hence the image classifier works less properly. Let us demonstrate it with an example. Imagine that the task is to recognize nine letters ('1', '2', '3', '4', '5', '6', '7', '8', '9'). If the letter '8' is rotated by 90 degree, the image classifier is unlikely to recognize that the augmented letter '∞' and the original letter '8' belong to the same category. The use of ±$\theta_R$ = ±180° raises the recognition performance back to 94% because the vertical symmetry almost reappears. As plotted in Figure 8, applying shear augmentation is destructive to recognize the "NFT" objects because the shape of nanowires (or fibers) is arguably rectangular. The shear matrix changes the shape of nanowires (or fibers) to parallelogram-like shapes and hence the software is hard to detect correctly. In view of this, the use of rotational augmentation and shear augmentation are not recommended for classifying 1D materials.

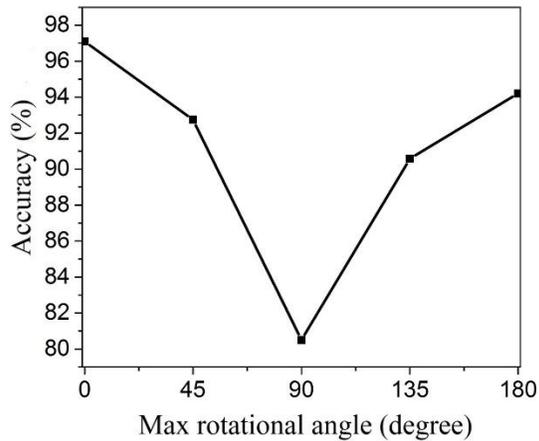

Figure 7: The "NFT" dataset is subjected to rotational augmentation.

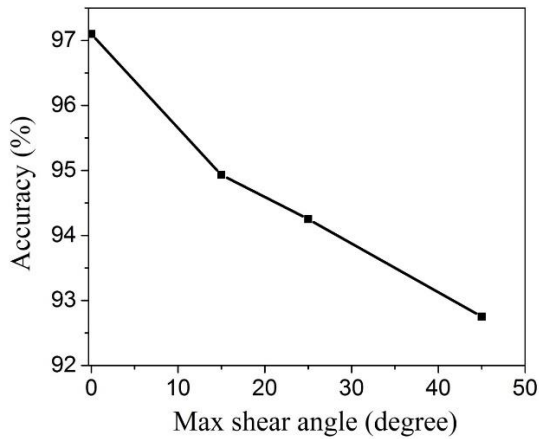

Figure 8: Shear augmentation is applied to the "NFT" dataset.

In contrast, rotational augmentation is very effective to increase the validation accuracy of the AAO images as plotted in Figure 9. The validation accuracy of the AAO images without the application of rotational augmentation is 86% only. However, the validation accuracy reaches 92% after setting $\pm\theta_R = \pm 90°$ or $\pm 135°$. If the reflection factor is applied, the validation accuracy of the AAO images is decreased by ~10% regardless of the value of $\theta_R$. Due to this empirical result, no validation accuracies in Figure 9 involve the reflection factor. Unlike the situation of the "NFT" objects, the holes in the AAO images under rotational augmentation do not break the vertical symmetry severely [25]. An analogy is that the circular shape of an orange is still maintainable even though the orange is rotated. As the geometric symmetry of the AAO holes is protected under rotational effect, applying the rotational augmentation creates a more valuable dataset for image classification that boost the validation accuracy to 92%. According to Figure 10, the shear matrix also provides a negative effect on classifying the AAO images because the shape of the AAO holes may be transformed from circle to ellipse (or vice versa) by the shear matrix that makes the image classification worse. Despite the scale factor improves the validation accuracy of the "NFT" objects, it is not recommended to recognize the porous materials with the evidence of Figure 11. Let's recall the fact that the size of the AAO holes depends on the applied voltage. Scaling the AAO images is equivalent to resizing the AAO holes virtually [28] and hence the use of scale factor augmentation makes the image classification very poor. Applying translational matrices do not enhance the validation accuracy of the AAO images even further.

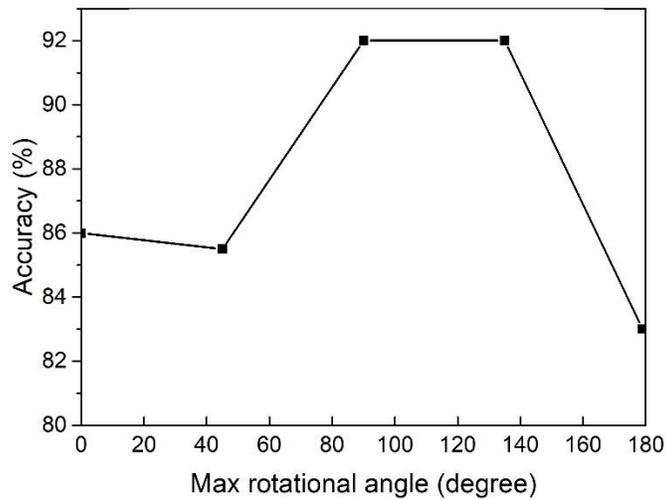

Figure 9: The validation accuracies of the AAO images under rotational augmentation.

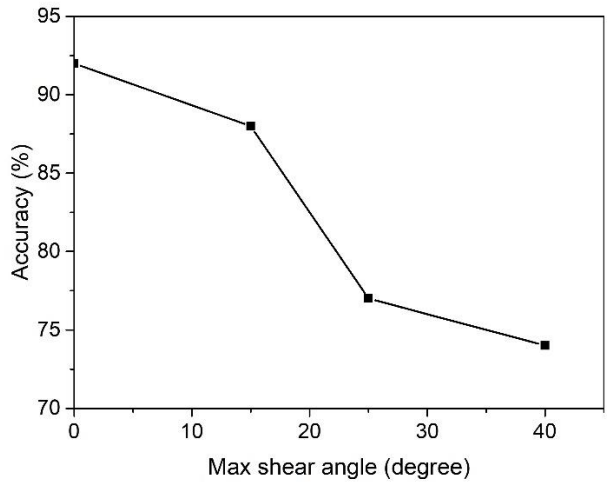

Figure 10: The shear augmentation drops the validation accuracy of the AAO images.

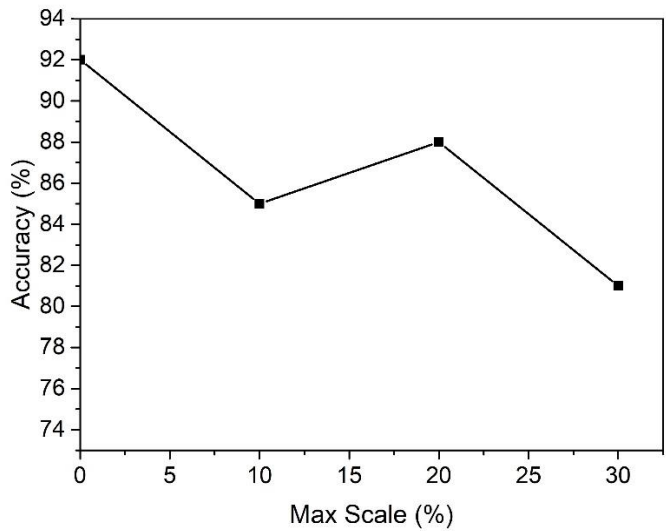

Figure 11: The validation accuracy of the AAO images as a function of the maximum scale factors.

Despite the total number of images per category is small in our dataset, we aim at understanding how data augmentations improve image classification instead of building a reliable pre-trained model at this stage [28]. By setting 20 images per category, we have already found that the rotational augmentation can benefit the image recognition of the AAO images. If more AAO images per category are assigned, the value of validation accuracy and the optimized $\theta_R$ may change but the model should draw the same conclusion again: i.e. applying the rotational augmentation improves the image classification of the AAO images. The prediction generated from a small dataset is arguably representative depending on which algorithm researchers use [29,30]. On the other hand, the classification of the "NFT" images is important for the future development in estimating the percentage of nanowires in nanowire-fiber mixtures. Every image-classification comes with a score point. Assume the image classification between nanowires and fibers are 100% in the two-class pre-trained model ideally, when the sample contains nanowires only, the score point of the SEM image is 100% nanowires vs 0% fibers. If the SEM image of a nanowire-fiber composite is analyzed by the two-class pre-trained model, the score point of the SEM image may be 60% nanowires vs 40% fibers where we may conclude that the nanowire-fiber composite contains 60% nanowires and 40% fibers. The same argument applies to the dissimilar AAO holes. For example, a two-class predictor is trained to identify the AAO substrate under 170V and 190V. After combining Figure 2d and Figure 2e in equal partition, the software shows that the combined figure belongs to the category of "170V" with the score point of 58%. It implies that the combined figure contains 58% AAO holes under the applied voltage of 170V (or 42% AAO holes are generated by the applied voltage of 190V). When scientists create the AAO holes in the radius of 5nm and 10 nm for embedding dissimilar nanowires, the score point of the SEM image will inform the probability of having AAO holes in the radius of 5nm. This score point will assist scientists to find out the composition in the nanowire-fiber mixture and the distribution of the AAO holes very quickly that presumably actuates the progress of materials science research.

## Conclusions:

By utilizing the GoogleNet network, we have found the desired types of data augmentations for classifying nanowires, fibers, tips and the porosity of anodized aluminum oxide (AAO). The line of symmetry plays the most dominant role to classify the above objects. These results assists the artificial-intelligence software to 'visualize' the composition of nanowire-fiber mixtures and the distribution of the AAO holes significantly. This work actuates the development of the artificial-intelligence software for the Nanoscience Foundries & Fine Analysis (NFFA) Europe Projects.

## Data availability statement

The raw/processed data required to reproduce these findings cannot be shared at this time due to technical or time limitations.

## References:

[1] Barret Zoph, Vijay Vasudevan, Jonathon Shlens, Quoc V. Le; Proceedings of the IEEE Conference on Computer Vision and Pattern Recognition, pp. 8697-8710 (2018)

[2] Li-Jia Li, Hao Su, Yongwhan Lim, Li Fei-Fei, Object Bank: An Object-Level Image Representation for High-Level Visual Recognition, International Journal of Computer Vision, Vol 107, pp 20–39 (2014)

[3] Xiuling Zhang, Kailun Wei, Xuenan Kang, Jinxiang Li, Hybrid nonlinear convolution filters for image recognition, Applied Intelligence, Vol 51, pp 980–990 (2021)

[4] Diego Inácio Patrício, Rafael Rieder, Computer vision and artificial intelligence in precision agriculture for grain crops: A systematic review, Computers and Electronics in Agriculture, Vol 153, pp 69-81 (2018)

[5] Niall O'Mahony, Sean Campbell, Anderson Carvalho, Suman Harapanahalli, Gustavo Velasco Hernandez, Lenka Krpalkova, Daniel Riordan, Joseph Walsh, Deep Learning vs. Traditional Computer Vision, Advances in Computer Vision, Vol 943, pp 128-144 (2019)

[6] Jiaoping Zhang, Hsiang Sing Naik, Teshale Assefa, Soumik Sarkar, R. V. Chowda Reddy, Arti Singh, Baskar Ganapathysubramanian & Asheesh K. Singh, Computer vision and machine learning for robust phenotyping in genome-wide studies, Scientific Reports, Vol 7, Article number: 44048 (2017).

[7] Fabian Horst, Sebastian Lapuschkin, Wojciech Samek, Klaus-Robert Müller & Wolfgang I. Schöllhorn, Explaining the unique nature of individual gait patterns with deep learning, Scientific Reports, Vol 9, Article number: 2391 (2019)

[8] Haiguang Wen, Junxing Shi, Yizhen Zhang, Kun-Han Lu, Jiayue Cao, Zhongming Liu, Neural Encoding and Decoding with Deep Learning for Dynamic Natural Vision, Cerebral Cortex, Vol 28, Issue 12, pp 4136–4160 (2017)

[9] Ajeet Ram Pathak, Manjusha Pandey, Siddharth Rautaray, Deep Learning Approaches for Detecting Objects from Images: A Review, Progress in Computing, Analytics and Networking. Advances in Intelligent Systems and Computing, Springer, Vol 710, pp 491-499 (2018)

[10] Muhammad Sajjad, Salman Khan, Khan Muhamm, WanqingWu, AminUllah, Sung Wook Baik, Multi-grade brain tumor classification using deep CNN with extensive data augmentation, Journal of Computational Science, Vol 30, pp 174-182 (2019)

[11] Hai Huang, Hao Zhou, Xu Yang, Lu Zhang, Lu Qi, Ai-Yun Zang, Faster R-CNN for marine organisms detection and recognition using data augmentation, Neurocomputing, Vol 337, pp 372-384 (2019)

[12] Dongmei Han, Qigang Liu, Weiguo Fan, A new image classification method using CNN transfer learning and web data augmentation, Expert Systems with Applications, Vol 95, pp 43-56 (2018)

[13] Zohaib Mushtaq, Shun-FengSu Quoc-Viet Tran, Spectral images based environmental sound classification using CNN with meaningful data augmentation, Applied Acoustics, Vol 172, 107581 (2021)

[14] Mohammad Hadi Modarres, Rossella Aversa, Stefano Cozzini, Regina Ciancio, Angelo Leto & Giuseppe Piero Brandino, Neural Network for Nanoscience Scanning Electron Microscope Image Recognition, Scientific Reports, Vol 7, Article number: 13282 (2017)

[15] NFFA-EUROPE. Draft metadata standard for nanoscience data. NFFA project deliverable D11.2, http://www.nffa.eu/media/124786/d112-draft-metadata-standard-for-nanoscience-data_20160225-v1.pdf.

[16] Eldar M. Khabushev, Dmitry V. Krasnikov, Orysia T. Zaremba, Alexey P. Tsapenko, Anastasia E. Goldt, and Albert G. Nasibulin, Machine Learning for Tailoring Optoelectronic Properties of Single-Walled Carbon Nanotube Films, J. Phys. Chem. Lett. 10, 21, pp 6962–6966 (2019)

[17] Chenxi Zhai, Tianjiao Li, Haoyuan Shi and Jingjie Yeo, Discovery and design of soft polymeric bio-inspired materials with multiscale simulations and artificial intelligence, J. Mater. Chem. B, 8, pp 6562-6587 (2020)

[18] Fangyi Guan, Yu Xie, Hanxiang Wu, Yuan Meng, Ye Shi, Meng Gao, Ziyang Zhang, Shiyan Chen, Ye Chen, Huaping Wang, and Qibing Pei, Silver Nanowire–Bacterial Cellulose Composite Fiber-Based Sensor for Highly Sensitive Detection of Pressure and Proximity, ACS Nano, 14, 11, pp 15428–15439 (2020)

[19] Feiyan Zhu, Leilei Zhang*, Kejie Guan, Hejun Li, and Yong Yang, Carbon Fiber Composites Containing Strongly Coupled $Si_3N_4$ Nanowire-Carbon Nanotube Networks for Aerospace Engineering, ACS Appl. Nano Mater. 3, 6, pp 5252–5259 (2020)

[20] Sheung Mei Ng, Hon Fai Wong, Hon Kit Lau, and Chi Wah Leung, Large-Area Anodized Alumina Nanopore Arrays Assisted by Soft Ultraviolet Nanoimprint Prepatterning, J. Nanosci. Nanotechnol. 12, 8, pp 6315 (2012)

[21] Sheung Mei Ng, Jian Zhuo Xin, Hon Fai Wong, Hon Kit Lau, Hai Tao Huang, and Chi Wah Leung, Hierarchical Nanoporous Alumina by Soft Ultraviolet Nanoimprint Prepatterning-Assisted Anodization, J. Nanoeng. Nanomanuf. 3, 2, 126-130 (2013)

[22] Rikiya Yamashita, Mizuho Nishio, Richard Kinh Gian Do & Kaori Togash, Convolutional neural networks: an overview and application in radiology, Insights into Imaging, Vol 9, pp 611–629 (2018)

[23] Luyl-Da Quach, Nghi Pham Quoc, Nhien Huynh Thi, Duc Chung Tran, Mohd Fadzil Hassan, Using SURF to Improve ResNet-50 Model for Poultry Disease Recognition Algorithm, 2020 International Conference on Computational Intelligence, pp. 317-321 (2020)

[24] Pengjie Tang, Hanli Wang, Sam Kwong, G-MS2F: GoogLeNet based multi-stage feature fusion of deep CNN for scene recognition, Neurocomputing, Vol 225, pp 188-197 (2017)

[25] Iacopo Masi, Anh Tuấn Trần, Tal Hassner, Gozde Sahin & Gérard Medioni, Face-Specific Data Augmentation for Unconstrained Face Recognition, International Journal of Computer Vision, Vol 127, pp 642–66 (2019)

[26] Jerubbaal John Luke, Rajkumar Joseph, Mahesh Balaji, Impact of image size on accuracy and generation of convolution neural networks, International Journal of Research and Analytical Reviews, Vol 6, Issue 1, pp 70-80 (2019)

[27] Emmanuel Okafor, Lambert Schomaker & Marco A. Wiering, An analysis of rotation matrix and colour constancy data augmentation in classifying images of animals, Journal of Information and Telecommunication, , 2:4, pp 465-491 (2019)

[28] Connor Shorten & Taghi M. Khoshgoftaar, A survey on Image Data Augmentation for Deep Learning, Journal of Big Data, Vol 6, Article number: 60 (2019)

[29] Shuo Feng, Huiyu Zhou, Hongbiao Dong, Using deep neural network with small dataset to predict material defects, Materials & Design, Vol 162, pp 300-310 (2019)

[30] Jostein Barry-Straume, Adam Tschannen, Daniel W. Engels, Edward Fine, An evaluation of training size impact on validation accuracy for optimized convolutional neural networks, SMU Data Science Review, Vol. 1, No. 4, Art. 12 (2018)